\pdfoutput=1
\documentclass[usenatbib]{mn2e}
\usepackage[implicit=false,breaklinks,colorlinks,citecolor=blue]{hyperref}
\usepackage{apjfonts}
\usepackage{amssymb}
\usepackage{bm}
\usepackage{amsmath}
\usepackage{ctable}
\usepackage{url}
\usepackage{breakurl}
\usepackage{fixltx2e} 

\newcommand{\be}{\begin{equation}}
\newcommand{\ee}{\end{equation}}

\newcommand{\paperone}{Paper {\small I}}

\newcommand{\gizmourl}{\burl{http://www.tapir.caltech.edu/~phopkins/Site/GIZMO.html}}

\newcommand{\vspacerpostplot}{\vspace{-0.4cm}}
\newcommand{\grainsize}{a_{d}}
\newcommand{\sizeparam}{\alpha}
\newcommand{\Lbox}{L_{\rm box}}

\newcommand\plotonesize[2]
{\centering \leavevmode \includegraphics[width={#2\columnwidth}]{#1}}
\newcommand{\plotsidesize}[2]
{\centering \leavevmode \includegraphics[width={#2\textwidth}]{#1}}
\newcommand{\acknowledgments}{\begin{small}\section*{Acknowledgments}\end{small}}
\newcommand\altaffilmark[1]{$^{#1}$}
\newcommand\altaffiltext[1]{$^{#1}$}
\title[Charged Dust Dynamics in GMCs]{The Dynamics of Charged Dust in Magnetized Molecular Clouds\vspace{-0.5cm}}

\author[Lee et al.]{
\parbox[t]{\textwidth}{ 
	Hyunseok Lee\altaffilmark{1}\thanks{E-mail:hlee2@mit.edu}, Philip F. Hopkins\altaffilmark{1}, \&\ Jonathan Squire\altaffilmark{1,2}
} 
\vspace*{6pt} \\
\altaffiltext{1}{TAPIR, Mailcode 350-17, California Institute of Technology, Pasadena, CA 91125, USA} \\
\altaffiltext{2}{Walter Burke Institute for Theoretical Physics, Pasadena, CA 91125, USA\vspace{-0.3cm}}
}

\date{Submitted to MNRAS, December, 2016\vspace{-0.6cm}}
\begin{document}
\maketitle
\label{firstpage}

\begin{abstract}

We study the dynamics of large, charged dust grains in turbulent giant molecular clouds (GMCs). Massive dust grains behave as aerodynamic particles in primarily neutral dense gas, and thus are able to produce dramatic small-scale fluctuations in the dust-to-gas ratio. \citet{hopkins.lee} directly simulated the dynamics of neutral dust grains in super-sonic MHD turbulence, typical of GMCs, and showed that the dust-to-gas fluctuations can exceed factor $\sim1000$ on small scales, with important implications for star formation, stellar abundances, and dust behavior and growth. However, even in primarily neutral gas in GMCs, dust grains are negatively charged and Lorentz forces are non-negligible. Therefore, we extend our previous study by including the effects of Lorentz forces on charged grains (in addition to drag). For small charged grains (sizes $\ll 0.1\,\mu m$), Lorentz forces suppress dust-to-gas ratio fluctuations, while for large grains (sizes $\gtrsim 1\,\mu m$), Lorentz forces have essentially no effect, trends that are well explained with a simple theory of dust magnetization. In some special intermediate cases, Lorentz forces can enhance dust-gas segregation. Regardless, for the physically expected scaling of dust charge with grain size, we find the most important effects depend on grain size (via the drag equation) with Lorentz forces/charge as a second-order correction. We show that the dynamics we consider are determined by three dimensionless numbers in the limit of weak background magnetic fields: the turbulent Mach number, a dust drag parameter (proportional to grain size) and a dust Lorentz parameter (proportional to grain charge); these allow us to generalize our simulations to a wide range of conditions.
	
\end{abstract}

\begin{keywords}
galaxies: formation --- 
star formation: general --- cosmology: theory --- planets and satellites: formation --- accretion, accretion disks --- instabilities --- turbulence
\vspace{-1.0cm}
\end{keywords}

\vspace{-1.1cm}
\section{Introduction}
\label{sec:intro}

Dust is crucial for diverse array of phenomena in astrophysics. Dust plays an important direct role in planet and star formation, and also in ``feedback'' processes during star cluster and galaxy formation. Dust is also vital for radiative cooling of gas, attenuation and absorption of light in the interstellar medium (ISM), and the evolution of heavy-element abundances and phases in galaxies. It is also a key observational tracer of the ISM in nearby regions and high-redshift galaxies. This broad importance of dust means that it is critical to understand dust dynamics, and grain clustering, in the ISM and star-forming regions.

It has long been known that dust grains do not necessarily move with gas in astrophysical fluids. In proto-planetary disks, in particular, a wide variety of conditions have been identified where different fluid conditions can produce orders-of-magnitude variations in the local dust-to-gas ratio, including: ``pressure'' traps, local ``vortex traps'' or ``turbulent concentration'' in turbulent disks, the streaming instability, ``zonal flows'' in magnetically active disks, and more \citep[see e.g.][]{bracco:1999.keplerian.largescale.grain.density.sims,cuzzi:2001.grain.concentration.chondrules,youdin.goodman:2005.streaming.instability.derivation,johansen:2007.streaming.instab.sims,carballido:2008.large.grain.clustering.disk.sims,bai:2010.grain.streaming.sims.test,bai:2010.grain.streaming.vs.diskparams,pan:2011.grain.clustering.midstokes.sims,dittrich:2013.grain.clustering.mri.disk.sims,jalali:2013.streaming.instability.largescales,hopkins:2014.pebble.pile.formation}. Large fluctuations in the density of aerodynamic particles relative to gas have also long been observed in terrestrial turbulence \citep{squires:1991.grain.concentration.experiments,fessler:1994.grain.concentration.experiments,rouson:2001.grain.concentration.experiment,gualtieri:2009.anisotropic.grain.clustering.experiments,monchaux:2010.grain.concentration.experiments.voronoi}. 

More recently, several studies have suggested that dust grains in GMCs or neutral galactic disks should exhibit similar fluctuations \citep{padoan:dust.fluct.taurus.vs.sims,hopkins:totally.metal.stars,hopkins.conroy.2015:metal.poor.star.abundances.dust} -- in terms of the aerodynamic drag equations, a grain of diameter $\sim 0.1-1\,\mu m$ in a typical GMC is analogous to a meter-sized boulder in a protoplanetary disk. And observations have identified small-scale ($\sim0.01-1\,$pc) fluctuations in the local dust-to-gas ratio of large grains in a number of nearby molecular clouds \citep{thoraval:1997.sub.0pt04pc.no.cloud.extinction.fluct.but.are.on.larger.scales,thoraval:1999.small.scale.dust.to.gas.density.fluctuations,abergel:2002.size.segregation.effects.seen.in.orion.small.dust.abundances,flagey:2009.taurus.large.small.to.large.dust.abundance.variations,boogert:2013.low.column.threshold.for.ice.mantle.formation}. Across different regions in the ISM, variations in extinction curves and emission/absorption features similarly suggest there may be large fluctuations in the relative abundance of large grains \citep{miville-deschenes:2002.large.fluct.in.small.grain.abundances,gordon:2003.large.variations.extinction.curves.in.lmc.smc.mw.sightlines,dobashi:2008.dust.gas.variations.observed.over.lmc.gmcs,paradis:2009.large.variation.across.lmc.in.dust.size.distrib}. The solar neighborhood, in particular, appears to exhibit an anomalous large-grain abundance \citep{kruger:2001.ulysses.large.grain.and.pioneer.compilation.data,frisch:2003.small.large.grains.decoupled.near.solar.neighborhood,meisel:2002.radar.micron.dust.ism.particles.many.large.grains,altobelli:2006.helios.large.local.interstellar.grains,altobelli:2007.cassini.confirms.large.microns.sized.interstellar.dust.grains,poppe:2010.new.horizons.confirms.ulysses.large.dust.measurements}. And \citet{hopkins:totally.metal.stars} suggested that this could explain some (but not all) of the variations in abundances observed within some star massive clusters.\footnote{Specifically, \citet{hopkins:totally.metal.stars} argue grain-gas dynamics may be relevant for certain abundance variations in large, low-density clusters (see e.g.\ \citealt{carretta:2009.gc.abundance.var.compilation,carraro:2013.ngc6791.review}), where the grain-decoupling parameter $\alpha$ (introduced below) is maximized, as opposed to low-mass clusters which appear to exhibit smaller abundance spreads (\citealt{pancino:2010.abundance.open.cluster,bragaglia:2012.berkeley.39.open.cluster.no.abundance.var,carrera:2013.open.cluster.spread.search}).}

But there are some important differences between dust dynamics in GMCs, as compared to the more well-studied terrestrial and planetary disk cases: most obviously, that the turbulence in GMCs is highly super-sonic, approximately isothermal (because the gas is rapidly-cooling), magnetized, and self-gravitating. \citet{hopkins.lee} (hereafter \paperone) presented a first numerical study of dust as aerodynamic particles under these conditions, and showed that indeed similar, dramatic fluctuations are expected in supersonic, isothermal, magnetohydrodnamic (MHD) turbulence, on scales that could be important for star formation, dust grain growth, and a wide variety of other phenomena. However, that study considered only neutral dust grains -- i.e.\ while the {\em gas} was magnetized, the grains felt no Lorentz forces. But real grains in GMCs are expected to be charged, and the Lorentz forces should dominate over aerodynamic (drag) forces for sufficiently small grains, or for large grains in sufficiently low-gas-density regions. This is yet another, perhaps critical, way that dust dynamics are different in GMCs and the ISM from terrestrial or proto-planetary disk turbulence.

In this paper, we therefore extend the study of \paperone, to include explicit, self-consistent Lorentz forces on grains, with realistic charges. We will show this does indeed have effects in the regimes expected, but these are generally sub-dominant to the effects of modest changes in the grain size.

\begin{figure*}
	\plotsidesize{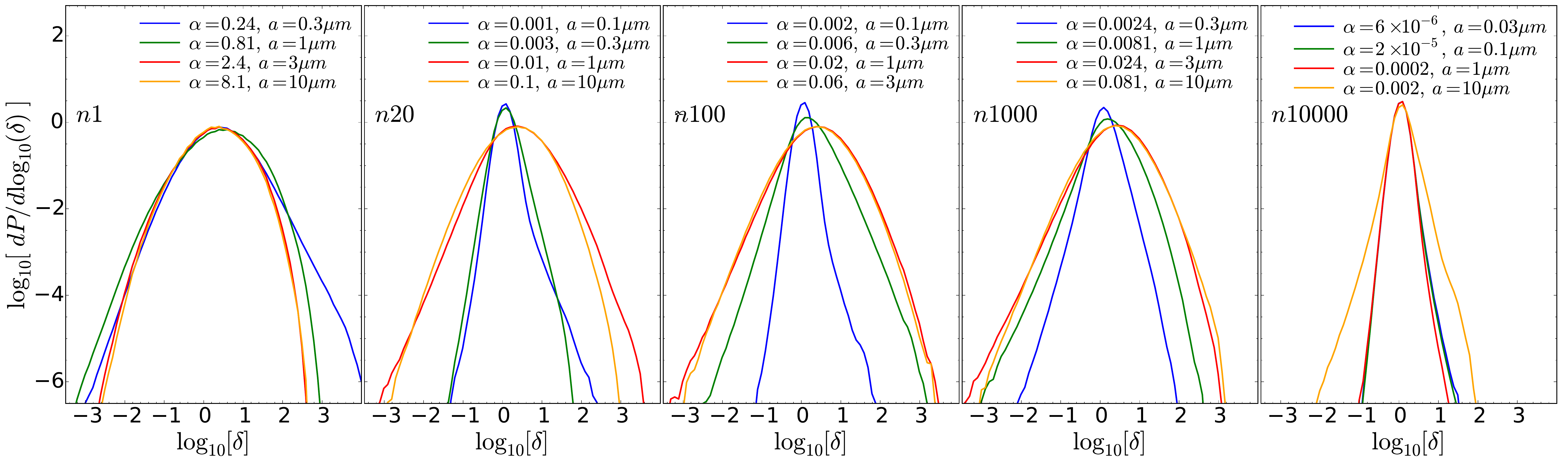}{1.001}
	\vspace{-0.7cm}
	\caption{Time-averaged distribution of dust-to-gas ratios $\delta\equiv (n_{\rm dust}/n_{\rm gas}) / (\langle n_{\rm dust} \rangle / \langle n_{\rm gas} \rangle)$ ($=1$ at mean densities; see Table~\ref{tbl:sims}) in the simulations (with ``normal'' Lorentz forces). Each frame shows a different box corresponding to a different mean physical density, and series of different grain drag/size parameters $\alpha$. As expected, $\alpha$ primarily governs the fluctuations: smaller $a_{d}$ and/or higher $\langle n_{\rm gas} \rangle$ reduce the variation $\delta$. As seen in \paperone, the distributions are approximately log-normal in their cores but exhibit strong power-law tails, with slope $dP/d\log\delta \propto \delta^{\pm2}$ (the steeper falloff at high-$\delta$ in the highest-$\alpha$ runs is closer to $\propto \delta^{-3}$). 
		\vspacerpostplot 
		\label{fig:vanilla}}
\end{figure*}

\begin{footnotesize}
\ctable[
caption={{\normalsize 3D Simulations of Dust-Gas Dynamics in This Paper}\label{tbl:sims}},center,star]{|c|ccccc|ccccc|ccc|}{
\tnote[ ]{List of simulations analyzed in the text: each is a 3D MHD simulation with driven isothermal turbulence, and grains of various sizes with dynamics determined by drag and Lorentz forces (Eq.~\ref{eqn:grain.eom}). Columns {\bf (2)}-{\bf (6)} show various dimensionless parameters for each simulation: $\mathcal{M}$, steady-state turbulent rms Mach number on the box scale (set by driving routines and mass-weighted); $\mathcal{M}_A$, steady-state Alfv\'enic Mach number $\mathcal{M}_A=\mathcal{M}/ v_A $ (where $v_A$ is the mass-weighted rms Alfv\'en speed); $\sizeparam$, grain drag/size parameter (Eq.~\ref{eqn:grain.size}); $\phi$, Lorentz parameter (Eq.~\ref{eqn:lorentz.phys}); $(\Theta_1,\Theta_2)$, dust magnetization parameters, which provide an estimate whether dust trajectories are significantly affected by the magnetic field (see Sec.~\ref{sec:scalings} and  Eq.~\ref{eqn: magnet params}).
{These dimensionless parameters allow us to study  a broad range of systems with fewer simulations and isolate the key controlling parameters. In other words, even with different physical parameters, two clouds
 at the same $\mathcal{M}$, $\mathcal{M}_A$, $\alpha$, and $\phi$ will exhibit the same physics.}
Columns {\bf (7)}-{\bf (10)} give an example of one specific set of physical cloud parameters which could produce the given $\mathcal{M}$, $\alpha$, $\phi$ -- here we select a typical isothermal temperature $T$ for cold molecular gas in local GMCs ($10\,$K) or warm gas in starburst regions ($70\,$K), a box size $L_{\rm box}$ and density $\langle n_{\rm box} \rangle$ roughly on the observed linewidth-size relation and size-mass relations from \citet{bolatto:2008.gmc.properties}, a set of physical (large) grain sizes $a_{d}$, and grain charges $Z_{d}$ determined from \citet{draine:1987.grain.charging} (see text). Columns {\bf (10)}-{\bf (13)} summarize the (time-averaged) dust-to-gas fluctuations measured from simulation in saturated steady-state: $\sigma_{\log \delta}$ is the logarithmic dispersion in $\delta$ ($\sigma_{\log \delta}^{2}\equiv \langle (\log_{10}\delta)^{2} \rangle - \langle \log_{10}\delta \rangle^{2}$) from Fig.~\ref{fig:vanilla}, where the local dust-to-gas ratio $\delta \equiv (n_{\rm dust}/n_{\rm gas}) / (\langle n_{\rm dust} \rangle /\langle n_{\rm gas} \rangle)$. $C\equiv \langle n_{\rm dust}^{2} \rangle / \langle n_{\rm dust} \rangle^{2}$ is the volume-weighted dust clumping factor. $C({\rm dense})$ is $C$ measured only in the dense ($n_{\rm gas} > \langle n_{\rm gas} \rangle$) gas. Note that clumping factors quoted in \paperone\ were accidentally weighted incorrectly, making them larger by a factor $\sim5$.
}
}{
\hline\hline
\multicolumn{1}{|c}{Simulation} &
\multicolumn{5}{|c}{Simulation Parameters} &
\multicolumn{5}{|c}{Example Physical Cloud Parameters} &
\multicolumn{3}{|c|}{Dust-Gas Fluctuations} \\
\multicolumn{1}{|c}{Name} &
\multicolumn{1}{|c}{$\mathcal{M}$} &
\multicolumn{1}{c}{$\mathcal{M}_A$} &
\multicolumn{1}{c}{$\sizeparam$} &
\multicolumn{1}{c}{$\phi$} &
\multicolumn{1}{c}{$(\Theta_{1},\Theta_{2})$} &
\multicolumn{1}{|c}{$\langle n\rangle (\rm cm^{-3})$} &
\multicolumn{1}{c}{$L(\rm pc)$} &
\multicolumn{1}{c}{$T(\rm K)$} &	
\multicolumn{1}{c}{$\grainsize(\rm \mu m)$} &
\multicolumn{1}{c}{$Z_{d}$} &
\multicolumn{1}{|c}{$\sigma_{\log\delta} (\rm dex)$} &
\multicolumn{1}{c}{$C$} &
\multicolumn{1}{c|}{$C({\rm dense})$} \\		
\hline
n1 & 6.5 & 0.6 & 0.24 & 2.2 &(0.1, 0.3)& 1 & 25 & 10 & 0.3 & -1 & 0.29 & 8.5 & 15\\
&  & & 0.81 & 0.29 &(1.9, 2.1)& & & & 1 & -1.5 & 0.36& 2.2 & 2.5 \\ 
&  & & 2.4 & 0.097 &(9.6, 6.1)& & & & 3 & -4.5 & 0.29 & 1.3 & 1.4 \\ 
&  & & 8.1 & 0.029 &(59, 21)& & & & 10 & -15 & 0.27& 1.1 & 1.1 \\ 
\hline
n20 & 17 & 1.5 & 0.001 & 4.4 &(0.01, 0.3)&20 & 100 & 10 & 0.1 & -1 & 0.024 & 7.4 & 9.7 \\
&  & & 0.003 & 0.48 &(0.2, 3.1)& & & & 0.3 & -1 & 0.039 & 7.3 & 9.7 \\ 
&  & & 0.01 & 0.07 &(2.1, 21)& & & & 1 & -1.7 & 0.25 & 12 & 17 \\ 
&  & & 0.1 & 0.007 &(65, 205)& & & & 10 & -17 & 0.26 & 3.7 & 6.3 \\ 
\hline
n20\_hiZ & 17 & 1.5 & 0.001 & 44 &(0.001, 0.03)& 20 & 100 & 100 & 0.1 & -10 & 0.035 & 7.9 & 10 \\
&  & & 0.003 & 4.8 &(0.02 0.3)& & & & 0.3 & -10 & 0.024 & 7.5 & 9.7 \\ 
&  & & 0.01 & 0.73 &(0.2, 2.1)& & & & 1 & -17 & 0.11 & 9.4 & 13 \\ 
&  & & 0.1 & 0.073 &(6.5, 20)& & & & 10 & -170 & 0.28 & 3.8 & 6.3 \\
\hline
n20\_m30 & 37 & 2.6 & 0.001 & 4.4 &(0.02, 0.6)& 20 & 100 & 10 & 0.1 & -1 & 0.019 & 8.4 & 11 \\
&  & & 0.003 & 0.48 &(0.3, 5.4)& & & & 0.3 & -1 & 0.032 & 7.8 & 11 \\ 
&  & & 0.01 & 0.073 &(3.6, 36)& & & & 1 & -1.7 & 0.16 & 10 & 15 \\ 
&  & & 0.10 & 0.0073 &(112, 356)& & & & 10 & -17 & 0.28 & 5.0 & 8.7 \\ 		 	
\hline
n20\_noZ & 17 & 1.5 & 0.0001 & 0 &$\infty$& 20 & 100 & 10 & 0.01 & 0 & 0.021 & 8.7 & 11 \\
&  & & 0.001 & 0 &$\infty$& & & & 0.1 & 0 & 0.053 & 8.9 & 12 \\ 
&  & & 0.01 & 0 &$\infty$& & & & 1 & 0 & 0.23 & 12 & 17 \\ 
&  & & 0.10 & 0 &$\infty$& & & & 10 & 0 & 0.30 & 4.0 & 7.1 \\ 		 	
\hline
n100 & 9.6 & 1.4 & 0.002 & 2.0 &(0.03, 0.7)& 100 & 10 & 10 & 0.1 & -1 & 0.020 & 6.2 & 8.1 \\
&  & & 0.006 & 0.22 &(0.5, 6.4)& & & & 0.3 & -1 & 0.12 & 7.9 & 11 \\ 
&  & & 0.02 & 0.033 &(6.0, 42)& & & & 1 & -1.7 & 0.26 & 11 & 16 \\ 
&  & & 0.06 & 0.011 &(31, 127)& & & & 3 & -5.0 & 0.27 & 7.0 & 11 \\ 		 	
\hline
n100\_noZ & 9.6 & 1.4 & 0.0002 & 0 &$\infty$& 100 & 10 & 10 & 0.01 & 0 & 0.019 & 6.6 & 8.4 \\
&  & & 0.002 & 0 &$\infty$& & & & 0.1 & 0 & 0.043 & 7.0 & 8.9 \\ 
&  & & 0.02 & 0 &$\infty$& & & & 1 & 0 & 0.22 & 13 & 17 \\ 
&  & & 0.20 & 0 &$\infty$& & & & 10 & 0 & 0.31 & 3.1 & 5.1 \\ 		 	
\hline
n1000 & 7.4 & 2.3 & 0.0024 & 0.069 &(1.6, 33)& 1000 & 2.5 & 10 & 0.3 & -1 & 0.042 & 5.1 & 6.6 \\
&  & & 0.0081 & 0.009 &(23, 255)& & & & 1 & -1.5 & 0.13 & 7.4 & 9.6 \\ 
&  & & 0.024 & 0.003 &(119, 767)& & & & 3 & -4.5 & 0.23 & 12 & 17 \\ 
&  & & 0.081 & 0.001 &(654, 2300)& & & & 10 & -15 & 0.25 & 7.8 & 12 \\ 
\hline		 	
n1000\_sm & 7.4 & 2.3 & 0.00024 & 6.9 &(0.005, 0.3)& 1000 & 2.5 & 10 & 0.03 & -1 & 0.017 & 6.1 & 7.9 \\ 
&  & & 0.0004 & 2.5 &(0.02, 0.9)& & & & 0.05 & -1 & 0.016 & 6.1 & 7.8 \\ 
&  & & 0.00081 & 0.62 &(0.1, 3.7)& & & & 0.10 & -1 & 0.016 & 6.1 & 7.8 \\ 
&  & & 0.0016 & 0.15 &(0.6, 15)& & & & 0.20 & -1 & 0.017 & 6.2 & 7.9 \\ 
\hline
n10000 & 18 & 2.8 & 0.000006 & 2.2  &(0.003, 1.3)& 10000 & 10 & 70 & 0.03 & -1 & 0.017 & 7.5 & 9.6 \\
&  & & 0.00002 & 0.20 &(0.6, 14)& & & & 0.1 & -1 & 0.017 & 7.3 & 9.4 \\ 
&  & & 0.0002 & 0.02 &(2.0, 140)& & & & 1 & -10 & 0.017 & 7.3 & 9.4 \\ 
&  & & 0.002 & 0.002 &(63, 1400)& & & & 10 & -100 & 0.032 & 7.4 & 9.6 \\ 		 	
\hline
n10000\_sz & 18 & 2.8 & 0.0002 & 0.02 &(2.0, 140)& 10000 & 10 & 70 & 1 & -10 & 0.020 & 7.9 & 9.5 \\
\hline\hline
}
\end{footnotesize}

\begin{figure}
	\plotonesize{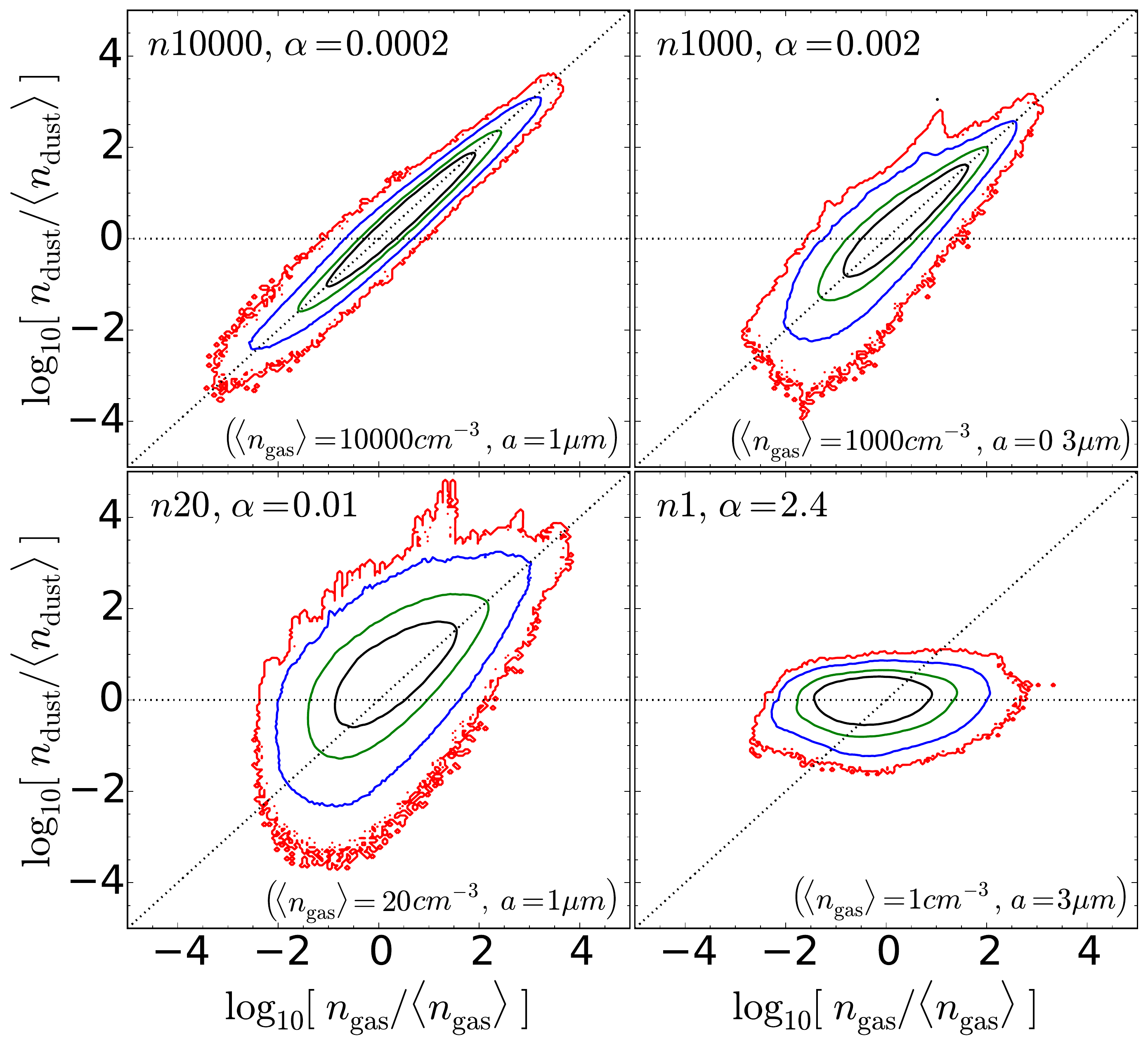}{0.97}
	\vspace{-0.25cm}
	\caption{Time-averaged bivariate distribution of dust and gas densities, in representative simulations from Table~\ref{tbl:sims}. We plot iso-density contours at fixed probability density levels $dP/d\log{n_{\rm gas}}\,d\log{n_{\rm dust}} = 10^{-1},\,10^{-2},\,10^{-4},\,10^{-7}$ (black, green, blue, red, respectively). Dotted lines show $n_{\rm dust}=\langle n_{\rm dust} \rangle$ (dust at constant density) and $n_{\rm dust} = (\langle n_{\rm dust} \rangle/\langle n_{\rm gas} \rangle)\,n_{\rm gas}$ (perfect dust-gas coupling, i.e.\ $\delta = 1$).  The simulations shown are chosen to reflect increasing value of the ``drag parameter'' $\alpha$ (Eq.~\ref{eqn:grain.size}), which determines how tightly-coupled the dust and gas are. As explained in Sec.~\ref{sec:scalings}, for $\alpha\ll \mathcal{M}^{-2}$, the coupling is near-perfect. For $\alpha\gtrsim1$, the dust is almost entirely un-correlated with the gas. Intermediate $\alpha$ show dust roughly following gas, $n_{\rm dust} \propto n_{\rm gas}$, but with large local fluctuations in $n_{\rm dust}$ at all $n_{\rm gas}$. 
    {Note that the distribution in n10000 is likely affected by  Poisson noise and may be even more tightly coupled to the gas than it appears: its distribution resembles the limiting $\alpha \ll 1$ case (where the 
dust is perfectly coupled to the gas) as shown in \paperone, Fig.~C1.}
		\vspacerpostplot 
		\label{fig:ndust.ngas}}
\end{figure}

\begin{figure}
	\plotonesize{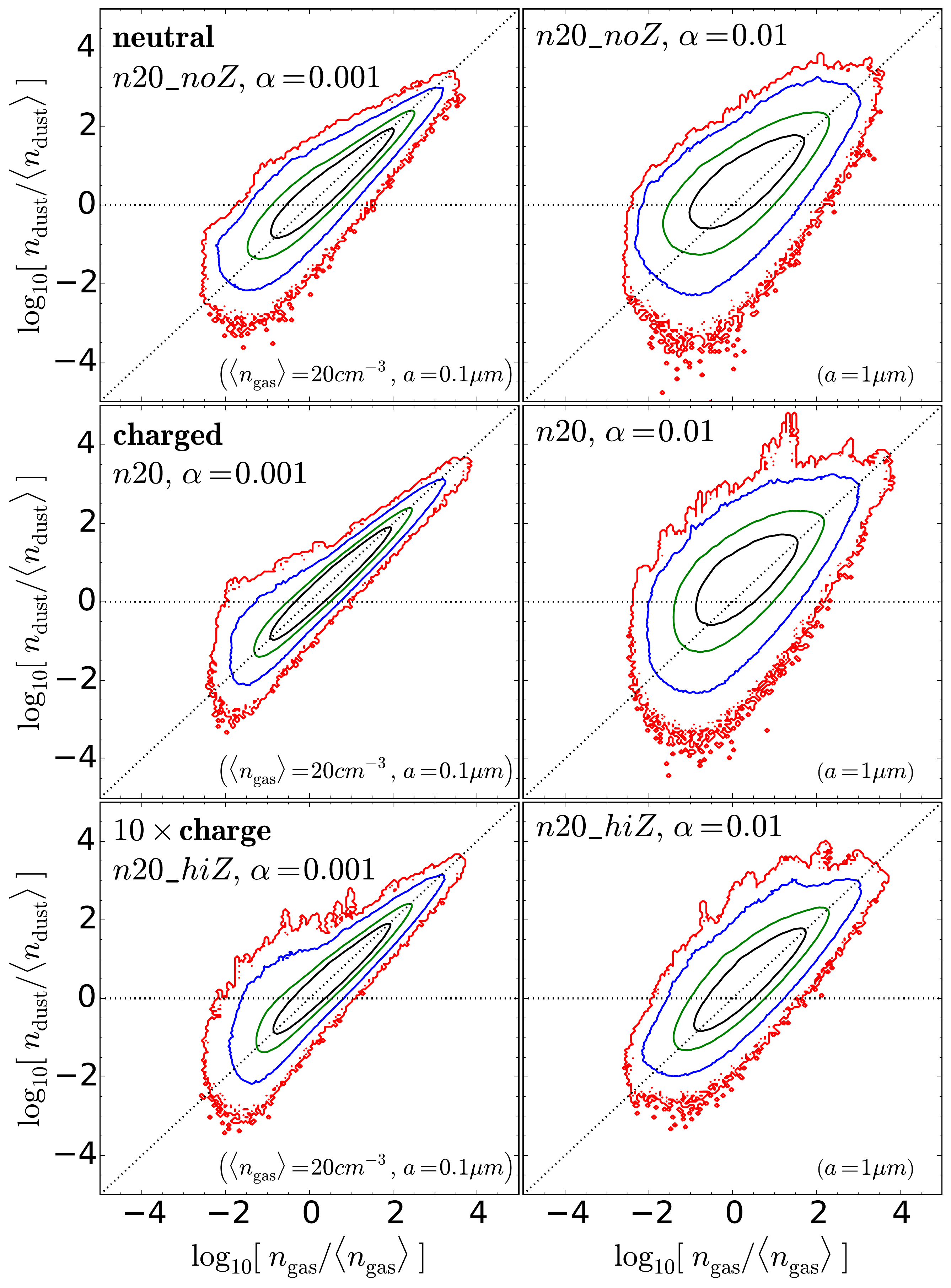}{0.97}
	\vspace{-0.25cm}
	\caption{
	Distribution of dust and gas densities (as Fig.~\ref{fig:ndust.ngas}), for two different grain sizes ($\alpha=0.001$, left column; $\alpha=0.01$, right column) and otherwise identical parameters. We compare cases with uncharged grains (``{n20\_noZ}''; top row), with ``standard'' grain charge  (``{n20}''; middle row), and with ``$10\times$'' grain charge (``n20\_hiZ''; bottom row); see Table~\ref{tbl:sims}. Lorentz forces reduce the fluctuations in $n_{\rm dust}(n_{\rm gas})$ for small grains, but have weak effects on large grains. This is explained by the dust-magnetization arguments outlined in Sec.~\ref{sec:scalings}. The small grains ($\alpha=0.001$) are in the magnetized regime ($\Theta_1<1$, $\Theta_2<1$) in both the ``n20'' and ``n20\_hiZ'' simulations,  and so are strongly affected by the magnetic field. In contrast, the larger grains ($\alpha=0.01$) are unmagnetized ($\Theta_1>1$) in ``n20'', so the dust-gas distribution is similar to that in ``n20\_noZ'' simualtion. The $\alpha=0.01$ grains in ``n20\_hiZ'' are in the ``mixed'' regime ($\Theta_1<1$, $\Theta_2>1$), which suggests that grains should be magnetized at low gas densities and unmagnetized at high gas densities. This idea is supported by the dust-gas distribution (bottom-right panel), which appears similar to that of ``n20'' and ``n20\_noZ'' at high gas densities, but shows comparatively reduced dust variance at low gas densities.
		\vspacerpostplot 
		\label{fig:ndust.ngas.lorentz}}
\end{figure}

\begin{figure}
	\plotonesize{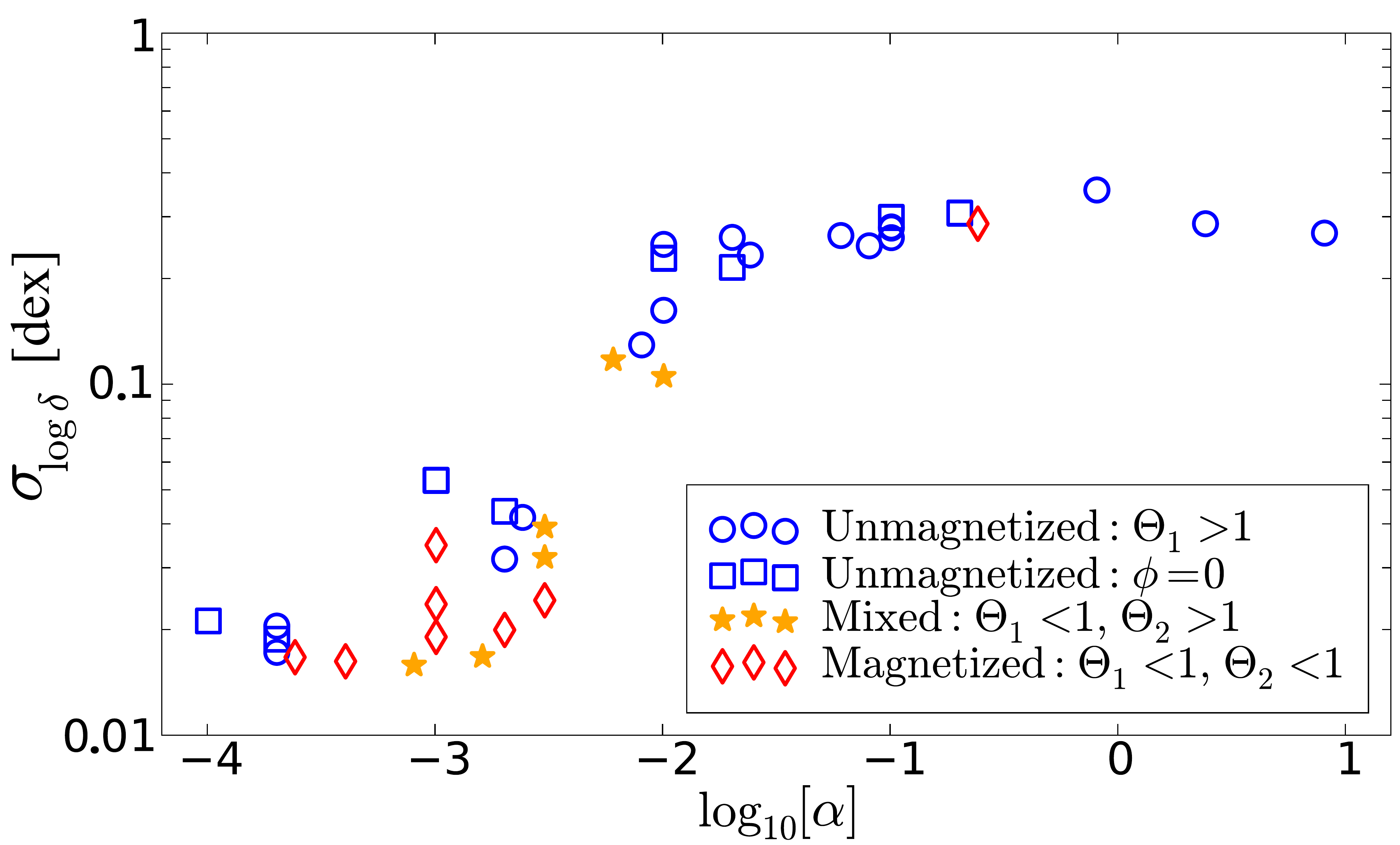}{0.99}
	\vspace{-0.25cm}
	\caption{
	Dispersion $\sigma_{\log \delta}$ of $\log_{10}(\delta)$, for all our simulations, as a function of $\alpha$ and magnetization  (see Eq.~\ref{eqn: magnet params}). Dust-to-gas fluctuations are primarily governed by $\alpha$, and $\sigma_{\log \delta}$ rises steeply from $<0.05$\,dex at $\alpha\lesssim 0.001$ to $\sim 0.3$\,dex at $\alpha\gtrsim0.1$; dust magnetization (and thus $\phi$) has only secondary effects. The saturation at high $\alpha$ is real, since the grains stop feeling drag forces, while the ``floor'' in $\sigma$ at $\sim 0.02$\,dex is a numerical artifact (see \paperone).
		\vspacerpostplot 
		\label{fig:2dparam}}
\end{figure}

\begin{figure}
\begin{tabular}{cc}
\hspace{-0.27cm}
\includegraphics[width=0.52\columnwidth]{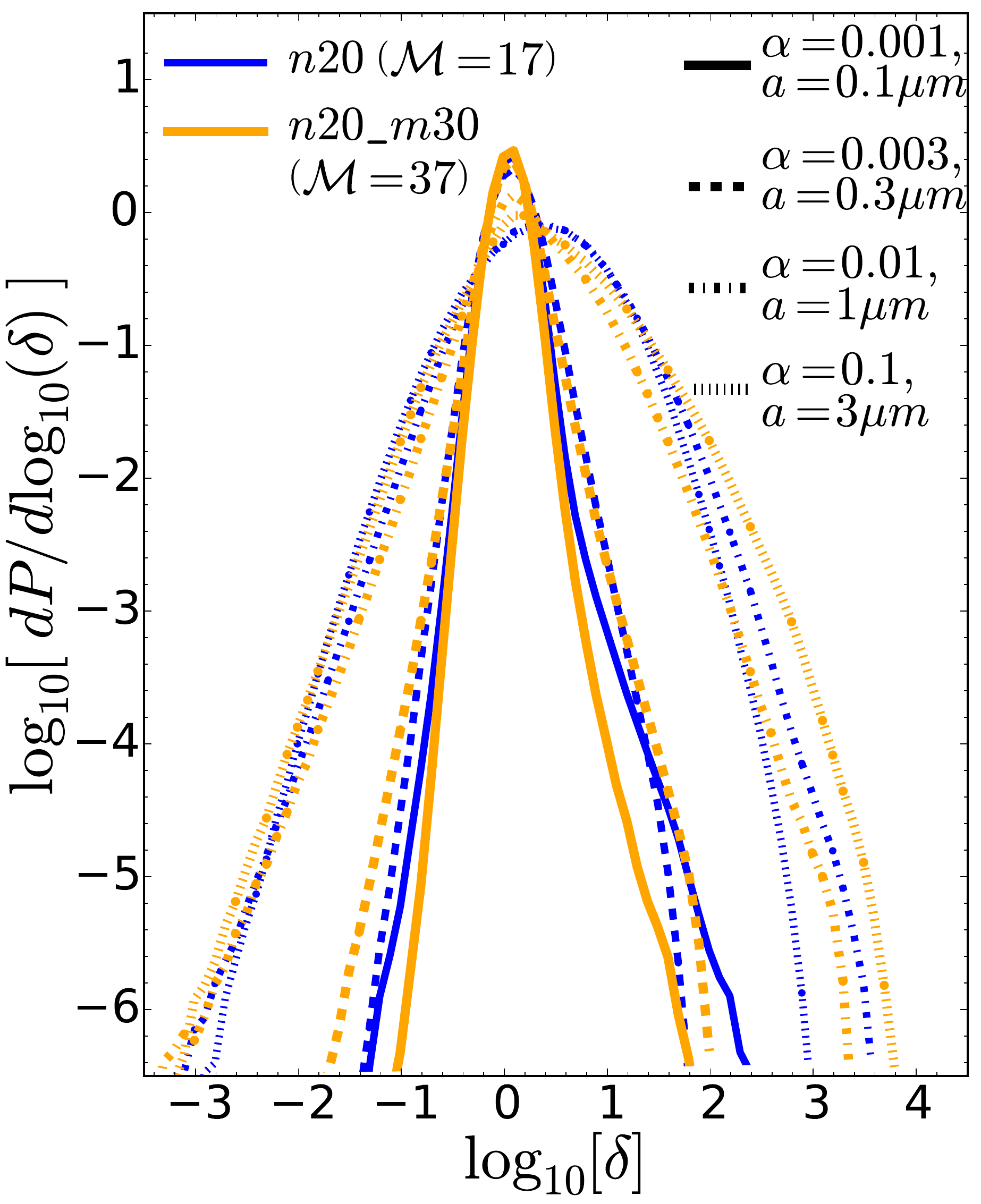} &
\hspace{-0.80cm}
\includegraphics[width=0.52\columnwidth]{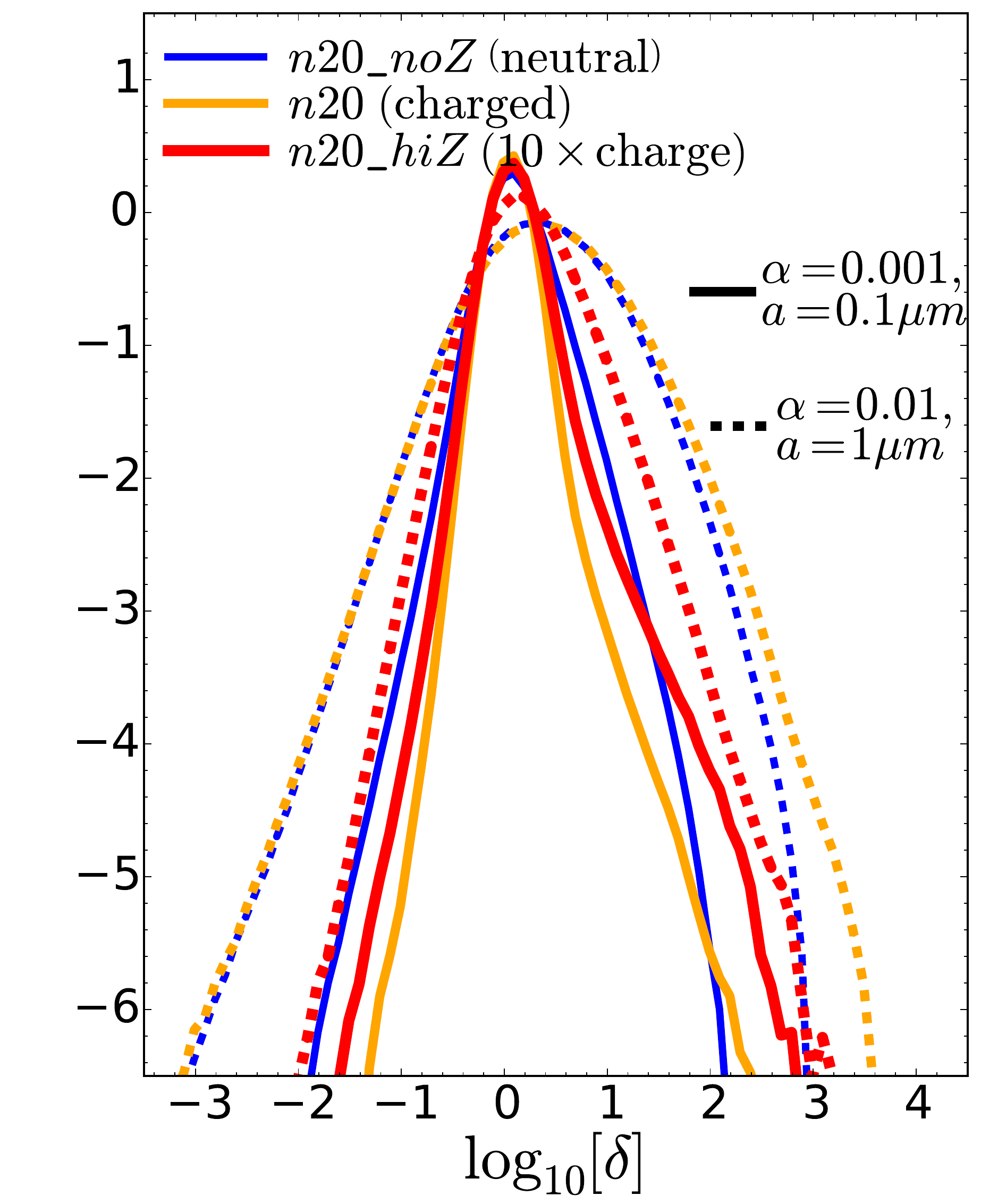} \\
\end{tabular}
	\vspace{-0.3cm}
	\caption{
	Distribution of $\delta$ (as Fig.~\ref{fig:vanilla}), for otherwise identical versions of the $\langle n_{\rm gas} \rangle=20\,{\rm cm^{-3}}$ box (``{n20}''), but varying either the Mach number ({left} panel) or grain charge/gas temperature ({right panel}). Increasing $\mathcal{M}$ from $\sim17$ to $\sim37$, we see weak effects: for all but the smallest grain sizes the distribution of $\delta$ is broadened, especially at the high-$\delta$ end, as the increased Mach number generates significantly larger variance in the {\em gas} density, hence more low-density regions where the grains are effectively de-coupled to the gas. The right panel shows the effect of adding grain Lorentz forces on the $\delta$ distribution. As expected from Fig.~\ref{fig:ndust.ngas.lorentz} (which shows the same simulations), going from neutral dust to normally charged dust (adding Lorentz forces) decreases the variance in $\delta$ for small grains and has a weak effect on large grains. Increasing the grain charge (and $\phi$) by a further factor of $10$ has a weak additional effect on the small grains (actually slightly {\em increasing} the variance in $\delta$),  but  significantly reduces the scatter in the large-grain case. This behavior is well explained by the dust-magnetization theory outlined in Sec.~\ref{sec:scalings}.
		\vspacerpostplot 
	\vspace{-1.5cm}
		\label{fig:machcomp.lorentzcomp}}
\end{figure}

\vspace{-0.5cm}
\section{Methods}
\label{sec:methods}

The methods here exactly follow \paperone\ with the addition of Lorentz forces, so we briefly summarize here and refer to that paper for details.  Our simulations use the code {\small GIZMO} \citep{hopkins:gizmo}.\footnote{A public version of this code is available at \gizmourl.} The gas obeys the equations of ideal magneto-hydrodynamics (MHD) with an isothermal equation of state, evolved using the Lagrangian ``MFM'' (meshless finite-mass) method; extensive tests of the method demonstrating excellent agreement with other well-studied higher-order codes on sub- and super-sonic MHD turbulence problems are presented in \citet{hopkins:gizmo,hopkins:mhd.gizmo}. This is solved in a 3D, periodic box with turbulence driven following \citet{bauer:2011.sph.vs.arepo.shocks}, as an Ornstein-Ulenbeck process with a specified turbulent Mach number and ``natural'' (equal) mix of compressive and solenoidal modes. The initial density is uniform and runs are initialized with a uniform magnetic field.  In  all runs except n1, this is taken to be small in comparison to the energy of the saturated turbulence; however, it is quickly amplified by field tangling and the {small-scale dynamo \citep{Brandenburg:2005kla}} to reach  approximate equipartition with the gas. The quoted values of Alfv\'enic Mach number $\mathcal{M}_A=\mathcal{M}/\langle v_A^2 \rangle^{1/2}$ (where $v_A = B/\sqrt{4\pi \rho}$) are measured from the saturated state, and increase with $\mathcal{M}$ as expected \citep{2011PhRvL.107k4504F}. All statistical quantities given in Table~\ref{tbl:sims} are measured after the turbulence has saturated.

Dust is represented as a population of tracer particles integrated on-the-fly through the fluid (representing grains of fixed size), with the equation of motion \citep[EOM;][]{draine.salpeter:ism.dust.dynamics}: 
\begin{align}
\label{eqn:grain.eom} \frac{d{\bf u}_{d}}{dt} &= -\frac{{\bf u}_{d}-{\bf u}_{\rm gas}}{t_{s}} + \frac{Z_{d}\,e}{m_{d}\,c} \,({\bf u}_{d}-{\bf u}_{\rm gas})\times{\bf B}\\ 
\label{eqn:tstop} t_{s} &\equiv \frac{\pi^{1/2}}{2\sqrt{2}}\,\left( \frac{\bar{\rho}_{d}\,\grainsize}{c_{s}\,\rho_{\rm gas}} \right)\,\left( 1 + \left|\frac{3\pi^{1/2}}{8\sqrt{2}}\,\frac{{\bf u}_{d}-{\bf u}_{\rm gas}}{c_{s}} \right|^{2}\right)^{-1/2}
\end{align}
where $d/dt$ is a Lagrangian derivative, ${\bf u}_{d}$, ${\bf u}_{\rm gas}$ are the grain and gas velocity, $c_{s}$ and $\rho_{\rm gas}$ the isothermal sound speed and density of the gas, $\bar{\rho}_{d}\sim2.4\,{\rm g\,cm^{-3}}$ is the internal (material) grain density \citep{draine:2003.dust.review}\footnote{Of course, the internal material properties of large dust grains, which may be complicated aggregates with ice mantles, are quite uncertain \citep[although see][who argue non-compactness would alter densities by only a small factor in ISM dust]{draine:2003.dust.review}. For this reason and others we will define the grain dynamics in terms of dimensionless quantities ($\alpha$ and $\phi$ below) which can then be trivially re-scaled for any material density.}, $\grainsize \sim 0.001-1\,\mu{\rm m}$ is the radius of single grain, $Z_{d}e$ the grain charge, $c$ the speed of light, and $m_{d}=(4\pi/3)\,\bar{\rho}_{d}\,a_{d}^{3}$ the grain mass. The first term is the drag term, with ``stopping time'' $t_{s}$, the second term is the Lorentz term in a magnetic field ${\bf B}$. The gas quantities are kernel-interpolated to the dust grain locations and the dust trajectories are integrated using a semi-implicit leapfrog scheme, as detailed in \paperone. In addition to {the Courant condition for neutral dust time-integration given in \paperone\ (Sec.~2.2),} stably integrating the Lorentz equation requires the dust timesteps be less than a small fraction (we take $10\%$) of the Larmor time $m_{d}\,c/(Z_{d}\,e\,|{\bf B}|)$. 

Radiation pressure, {gas self-gravity,} dust-dust collisions, and back-reaction of dust grains on gas are not included here:  {arguments in Sec.~2.4 of \paperone\ show} these all tend to be highly sub-dominant to drag and Lorentz forces in the systems we simulate here (in clouds of relatively low metallicity). Introducing them also breaks the scale-free nature of the problem here, necessitating a broader simulation survey and a different type of simulation (following star formation self-consistently, for example). {However, such effects can be important in 
specific situations (e.g., dust back reaction on the gas in the highest-density regions,  radiation pressure on dust near massive stars) and in future work we will examine some of these effects 
in more detail.}

Without loss of generality, we work in code units where $c_{s}=1$, the box length $L_{\rm box}=1$, and the box gas mass $M_{\rm box}\equiv \langle \rho_{\rm gas} \rangle \,L_{\rm box}^{3} = 1$; Eq.~\ref{eqn:grain.eom} then becomes 
\begin{align}
\alpha\,\frac{d \tilde{\bf u}_{d}}{d \tilde{t}} &= -\Delta\tilde{\bf u}\,\tilde{\rho}_{\rm gas}\,f(\Delta \tilde{\bf u}) + \phi\,\Delta \tilde{\bf u}\times\tilde{\bf B} \\ 
\Delta \tilde{\bf u} &\equiv \tilde{\bf u}_{d} - \tilde{\bf u}_{g} \ \ \ , \ \ \ f^{2} \equiv \frac{8}{\pi} + \frac{9}{16}\,|\Delta \tilde{\bf u}|^{2} \\ 
\label{eqn:grain.size}\alpha \equiv& \frac{\bar{\rho}_{d}\,a_{d}}{\langle \rho_{\rm gas} \rangle\,L_{\rm box}} \approx 20\,\frac{a_{d}}{\mu m}\,\frac{{\rm pc}}{\Lbox}\, \frac{{\rm cm^{-3}}}{\langle n_{\rm gas} \rangle}\,\frac{\bar{\rho}_{d}}{2.4\,{\rm g\,cm^{-3}}} \\ 
\label{eqn:lorentz.phys} \phi \equiv& \frac{3\,Z_{d}\,e}{4\pi\,c\,a_{d}^{2}\,\langle \rho_{\rm gas} \rangle^{1/2}}
\approx 0.2\,Z_{d}\,\left( \frac{\mu m}{a_{d}} \right)^{2} \left(\frac{\rm cm^{-3}}{\langle n_{\rm gas} \rangle} \right)^{1/2}
\end{align}
where the tilde-superscript $\tilde{x}$ denotes the value of $x$ in code units.\footnote{We will, for convenience, define the number density $\langle n_{\rm gas} \rangle  \equiv \langle \rho_{\rm gas} \rangle / \mu\,m_{p}$ with $\mu=2.3$ (appropriate for solar metallicity, molecular gas), but $\mu$ never enters the dynamics.} Here $\alpha$ is the dimensionless ``drag'' or ``size'' parameter from \paperone\ (smaller $\alpha$ means stronger gas-grain coupling), while the ``Lorentz parameter'' $\phi$ governs the effect of the Lorentz force.

In isothermal MHD turbulence starting from trace magnetic fields, the statistics of $\tilde{\bf u}_{g}$ and the saturated $\tilde{\bf B}$ are determined by the box-averaged Mach number $\mathcal{M} \equiv \langle |{\bf u}_{\rm gas}|^{2} \rangle^{1/2} / c_{s} = \langle |{\bf \tilde{u}}_{\rm gas}|^{2}\rangle^{1/2}$ (the statistics of the forcing can also be important; \citealt{Federrath:2013gu}). Thus, aside from the initial mean field (energetically important in the saturated state only for run ``n1''), the saturated dynamics of the problem are governed by $\mathcal{M}$, $\alpha$, and $\phi$, greatly simplifying our parameter survey. {All runs here use $256^{3}$ gas and $2\times256^{3}$ dust particles and a variety of convergence studies in \paperone\ (Appendix~B, Fig.~B1) demonstrate that this is sufficient for converged results in quantities studied here.} Table~\ref{tbl:sims} presents the list of simulations we study. For each run, we select $\mathcal{M}$, $\alpha$, $\phi$ by initially selecting a representative temperature $T$ for cold, molecular gas and box size $L_{\rm box}$, then estimating the Mach number and mean density corresponding to observed GMCs on the linewidth-size and mass-size relations from \citet{bolatto:2008.gmc.properties}. We then populate the box with four (relatively large)\footnote{We focus on large grains $a_{d}\sim0.1-10\,\mu m$ because (1) these contain most of the dust mass, and (2) smaller grains are tightly-coupled to the gas and therefore exhibit less extreme dust-to-gas fluctuations.} physical grain sizes ($\alpha$), and determine the charge $Z_{d}$ (and $\phi$) of each following \citet{draine:1987.grain.charging}\footnote{\citet{draine:1987.grain.charging} estimate grain charges based on a pure collisional model for large grains and a polarization model for small grains. \citet{shull:1978} show that accounting for higher-order effects can lower the charge by a factor $\sim2$ when the dust-gas motion is highly supersonic. Since this is uncertain we remind the reader that the parameter $\phi$ is what actually enters the dynamical equations solved here.}: $\langle Z_{d} \rangle = -1/(1+0.037\,\tau^{-1/2}) - 2.5\,\tau$ where $\tau\equiv a_{d}\,k\,T/e^{2}$. For comparison, we consider some cases with $Z_{d}=0$ (no Lorentz forces; our ``noZ'' runs) or $10\times$ larger charge (our ``hiZ'' runs) -- this crudely corresponds to the expected values if the gas were $10\times$ hotter (but retained the same Mach number). One run, ``n10000\_sz'' is run with twice as many gas elements and all dust grains the same size, to verify convergence; we find the statistics are nearly identical to our ``standard'' run.

\section{Dust magnetization}
\label{sec:scalings}

In this section, we estimate the influence of the magnetic field on dust-grain dynamics. A convenient way to parameterize this is through the ratio of the dust gyroradius $r_{\mathrm{gy},d}$ to its free-streaming length in the gas $L_{\mathrm{stream}}$. If $r_{\mathrm{gy},d}>L_{\mathrm{stream}}$, then the dust is effectively unmagnetized (it is stopped before undergoing a gyro-orbit), while if $r_{\mathrm{gy},d}<L_{\mathrm{stream}}$ the magnetic field will have a strong influence on  dust dynamics \citep{2002ApJ...566L.105L}. Here we estimate $r_{\mathrm{gy},d}/L_{\mathrm{stream}}$  assuming a basic supersonic turbulence model. Although qualitative, these arguments aid in understanding when dust charging should significantly modify its density distribution. As discussed below (see Sec.~\ref{sec:results}), we find reasonable agreement between the theory and simulation results.

The dust free-streaming length may be estimated as $L_{\mathrm{stream}}\sim \langle | {\bf u}_{d} - {\bf u}_{\mathrm{gas}}| \rangle \, t_{s}$, where the stopping time $t_{s}$ is given by Eq.~\eqref{eqn:tstop} and the relative velocity $\langle | {\bf u}_{d} - {\bf u}_{\mathrm{gas}}| \rangle$ may be estimated as the ``eddy velocity''  $v_{\lambda}\sim \langle |{\bf u}_{\mathrm{gas}}({\bf r}+{\bf \bm{\lambda}})- {\bf u}_{gas}({\bf r}) |\rangle$ (with $\lambda =|{\bf \bm{\lambda}}|$) of the turbulence on scale $\lambda = L_{\mathrm{stream}}$.\footnote{This is because velocities on smaller scales do not strongly perturb the dust, while those on larger scales simply advect  dust and gas together; see \citet{voelk:1980.grain.relative.velocity.calc,2002ApJ...566L.105L}; \paperone.} 
We then assume the standard hydrodynamic velocity scalings  $v_{\lambda}\sim \mathcal{M} c_{s}(\lambda/L_{\mathrm{box}})^{1/2}$ for $v_{\lambda} > c_{s}$; $v_{\lambda}\sim c_{s}(\lambda/R_{\mathrm{sonic}})^{1/3}$ for $v_{\lambda} < c_{s}$,  where $R_{\mathrm{sonic}}\sim L_{\mathrm{box}} \mathcal{M}^{-2}$ is the scale at which the turbulence transitions from supersonic to subsonic. This scaling  assumes that the influence of the magnetic field on the flow should be relatively unimportant until at, or below, the subsonic scales ($\lambda\leq R_{\mathrm{sonic}}$).\footnote{Note that we  have also assumed a subsonic scaling $v_{\lambda}\sim \lambda^{1/3}$, which only holds perpendicular to the magnetic field in magnetized turbulence \citep{Goldreich:1995hq,Maron:2001cs}. This estimate is more appropriate than the parallel scaling $v_{\lambda_{\parallel}}\sim \lambda_{\parallel}^{1/2}$ when the dust gyroradius is larger than the smallest perpendicular scales, since the dust will not be perfectly tied to the field lines (see \citealt{2002ApJ...566L.105L,Yan:2004vh} for further discussion). } Otherwise -- i.e.,  in the case of strong mean fields -- the turbulence would  be Alfv\'enic in character and anisotropic at  large scales \citep{Lithwick:2001iv,2003MNRAS.345..325C}. Our analysis is thus restricted to turbulence where $\mathcal{M}_{A}> 1$ and the field is tangled on supersonic scales (this is the opposite regime to \citealt{Yan:2004vh}). This appears to be  satisfied for most of the simulations detailed in Table~\ref{tbl:sims} (an exception is ``n1'', which has a relatively strong mean field).

Assuming the force on dust from the magnetic field will not strongly alter the streaming length,  one can estimate (see \paperone)
\begin{equation}
\frac{L_{\mathrm{stream}}}{L_{\mathrm{box}}}\sim  \begin{cases}
\alpha\left( \frac{n_{\mathrm{gas}}}{\langle n_{\mathrm{gas}}\rangle}\right)^{-1}  & L_{\mathrm{stream}}>R_{\mathrm{sonic}}\\[2ex]
\alpha^{3/2}\mathcal{M}\left( \frac{n_{\mathrm{gas}}}{\langle n_{\mathrm{gas}}\rangle}\right)^{-3/2}  & L_{\mathrm{stream}}<R_{\mathrm{sonic}}
\end{cases}.\label{eqn: Ls}\end{equation}
The transition between the two regimes occurs when ${n_{\mathrm{gas}}}/{\langle n_{\mathrm{gas}}\rangle}\sim \alpha \mathcal{M}^{2}$. Noting that the density contrast in an isothermal shock is  $n_{\mathrm{gas}}/\langle n_{\mathrm{gas}}\rangle \sim \mathcal{M}^{2}$ {\citep{Passot:1998cr,Konstandin:2012ei,konstantin:mach.compressive.relation}}, we see that there are three regimes (see \paperone, Sec.~3.1 for further discussion): (i) if $\alpha \gtrsim 1$, $L_{\mathrm{stream}}>R_{\mathrm{sonic}}$ everywhere (including the shocks) and the dust is  weakly coupled to the gas; (ii) if $\mathcal{M}^{-2}\lesssim \alpha\lesssim 1$ the dust is trapped in the highest density shocks but can cluster on scales larger than the sonic length; (iii) if $\alpha \ll \mathcal{M}^{-2}$ the dust is strongly coupled to the gas down to scales below the sonic length. 

To estimate the ratio $r_{\mathrm{gy},d}/L_{\mathrm{stream}}$, we assume $B/\rho_{\mathrm{gas}}^{1/2}\sim c_{s}\mathcal{M}/\mathcal{M}_{A}\sim \mathrm{constant}$, everywhere in the turbulence. This estimate is supported by observations \citep{1999ApJ...520..706C, 2010ApJ...725..466C} and numerical simulations \citep{Burkhart:2009dm,2009MNRAS.398.1082B,Molina:2012iv} in the regime of interest where the fields dynamically unimportant on supersonic scales (i.e., we again require $\mathcal{M}_{A}>1$).\footnote{\citet{2010ApJ...725..466C} report a lower density bound, below which the field and density are uncorrelated. This might be expected as the turbulence transitions into an Alfv\'enic regime, but we ignore this possible change in scaling here for simplicity.}
Note that the dust feels the total ``large-scale'' magnetic field (in contrast to the velocity field, where only $v_{\lambda}$ is important) so we do not need the magnetic field spectrum.  Using the dust gyroradius  $r_{\mathrm{gy},d}=m_{d} c |{\bf u}_{d}|/Z_{d}eB$ and Eqs.~\eqref{eqn:grain.size}--\eqref{eqn:lorentz.phys}, one obtains
\begin{equation}
\frac{r_{\mathrm{gy},d}}{L_{\mathrm{stream}}} \sim \frac{\alpha}{\phi} \left( \frac{n_{\mathrm{gas}}}{\langle n_{\mathrm{gas}}\rangle}\right)^{-1/2} \left(\frac{L_{\mathrm{stream}}}{L_{\mathrm{box}}}\right)^{-1} v_{L_{\mathrm{stream}}} \frac{\mathcal{M}_{A}}{\mathcal{M}}.\label{eqn: rgy}
\end{equation}
Inserting Eq.~\eqref{eqn: Ls} into Eq.~\eqref{eqn: rgy} leads to the estimate
\begin{equation}
\frac{r_{\mathrm{gy},d}}{L_{\mathrm{stream}}}\sim  \begin{cases}
\frac{\alpha^{1/2}}{\phi} \mathcal{M}_{A}  & L_{\mathrm{stream}}>R_{\mathrm{sonic}}\\[2ex]
\frac{1}{\phi}\left(\frac{{n_{\mathrm{gas}}}}{\langle n_{\mathrm{gas}}\rangle}\right)^{1/2} \frac{\mathcal{M}_{A}}{\mathcal{M}}  & L_{\mathrm{stream}}<R_{\mathrm{sonic}}
\end{cases}.\label{eqn: rgy over Ls}\end{equation}

Equation~\eqref{eqn: rgy over Ls} illustrates that the dust magnetization, ${r_{\mathrm{gy},d}}/{L_{\mathrm{stream}}}$, is governed by the parameters
\begin{equation}
\Theta_{1}=\frac{\alpha^{1/2}}{\phi}\mathcal{M}_{A}\quad \mathrm{and }\quad \Theta_{2} = \frac{1}{\phi}\mathcal{M}_{A}.\label{eqn: magnet params}\end{equation}
Recalling that the transition between the $L_{\mathrm{stream}}>R_{\mathrm{sonic}}$ and $L_{\mathrm{stream}}<R_{\mathrm{sonic}}$ regimes occurs at ${n_{\mathrm{gas}}}/{\langle n_{\mathrm{gas}}\rangle}\sim \alpha \mathcal{M}^{2}$, and noting that ${r_{\mathrm{gy},d}}/{L_{\mathrm{stream}}}$ increases monotonically with density, we see that there are three regimes:
\begin{description}
\item[${\bf \bm{\Theta_{1}} >1}$ \textbf{-- unmagnetized}:] {The dust is always ``unmagnetized'' ($r_{\mathrm{gy},d}>L_{\mathrm{stream}}$ over all scales).}
\item[${\bf \bm{\Theta_{1}} <1 \:\: \mathrm{\bf and} \:\: \bm{\Theta_{2}} >1}$ \textbf{-- mixed}:]{The dust is magnetized at low gas densities $n_{\mathrm{gas}}<n_{g,\mathrm{crit}}$, but switches to being unmagnetized ($r_{\mathrm{gy},d}>L_{\mathrm{stream}}$) as it streams into high  density regions $n_{\mathrm{gas}}>n_{g,\mathrm{crit}}$. The critical gas density that governs the change is
\begin{equation} 
\frac{n_{g,\mathrm{crit}}}{\langle n_{\mathrm{gas}}\rangle} \sim \phi^{2} \left( \frac{\mathcal{M}}{\mathcal{M}_{A}}\right)^{2}. \label{eq: critical density}\end{equation}
}
\item[${\bf \bm{\Theta_{1}} <1 \:\: \mathrm{\bf and} \:\: \bm{\Theta_{2}} <1}$ \textbf{-- magnetized}:] {The dust is always magnetized ($r_{\mathrm{gy},d}<L_{\mathrm{stream}}$ over all scales). This is because the density in the shocked regions ${n_{\mathrm{gas}}}/{\langle n_{\mathrm{gas}}\rangle}\sim \mathcal{M}^{2}$ is still not sufficiently high to make $r_{\mathrm{gy},d}>L_{\mathrm{stream}}$. }
\end{description}
The simulations presented below  cover each of these regimes (see $(\Theta_{1},\Theta_{2})$ column of  Table~\ref{tbl:sims}). Note that our discussion here has been intended to estimate \emph{when} the magnetic field is important for the dust, as opposed to the \emph{influence} of the magnetic field on the dust distribution (this 
 is discussed in more detail in the next section).


\vspace{-0.5cm}
\section{Results}
\label{sec:results}

In Figs.~\ref{fig:ndust.ngas}-\ref{fig:ndust.ngas.lorentz} we show the bivariate distribution of dust and gas densities in some representative simulations.\footnote{The dust and gas densities are determined in post-processing as in \paperone, using an adaptive kernel density estimator enclosing the nearest $\approx 64$ particles at the location of every dust particle to evaluate the dust density $n_{\rm dust}(a_{d})$ (counting species only of the same size) and $n_{\rm gas}$ at the same location. We have verified that the results are insensitive to the number of neighbors or shape of the kernel. After the first few turbulent crossing times the simulation reaches steady-state and there are no significant trends with time, so we simply average all snapshots together after this time.} Here and throughout this paper, all distribution functions are dust-mass weighted.
In Fig.~\ref{fig:ndust.ngas} we see the effects of increasing $\alpha$ (grain size). As described above (see Eq.~\ref{eqn: Ls}) and in \paperone, small grains ($\alpha \ll \mathcal{M}^{-2}$) are tightly coupled to gas, very large grains ($\alpha >1 $) are spread closer to uniformly and weakly-coupled to the gas, and grains with intermediate $\alpha$ ($\mathcal{M}^{-2} \lesssim \alpha \lesssim 1$) produce interesting dust-gas distributions.\footnote{Note, in Fig.~\ref{fig:ndust.ngas}, for the same grain size, the fluctuations at a given physical density (say, $n_{\rm gas}\sim10^{3}\,{\rm cm^{-3}}$), in the ``n20'' run (at $50\times$ mean density) are much larger than the fluctuations at the same mean density in the ``n1000'' or ``n10000'' run. This is discussed in \paperone\ and \citet{hopkins:totally.metal.stars,hopkins.conroy.2015:metal.poor.star.abundances.dust}. Basically, because the high-density regions in the lower-mean-density box form from a wide range of progenitor regions with lower pre-shock/compression density (hence weaker dust-gas coupling), their dust-to-gas ratio fluctuations are ``seeded'' in these progenitors. Once a region locally becomes sufficiently dense that grains are tightly trapped, these are ``frozen in,'' while new local fluctuations are suppressed.} This behavior
is similar to that seen in \paperone\ without Lorentz forces; for a more detailed analysis of the gas-density dependence of fluctuations, and their dependence on spatial scale (power spectra/correlation functions), we refer interested readers to \paperone. 

Figure~\ref{fig:ndust.ngas.lorentz} shows the effect of adding Lorentz forces at two grain sizes for three different levels of dust charge (no charge, ``n20\_noZ''; ``standard'' charge, ``n20''; and $10\times$ charge, ``n20\_hiZ''). The illustrated dust-gas distributions broadly follow our expectations based on the theory of dust magnetization in Sec.~\ref{sec:scalings}. Small grains ($\alpha=0.001$) in both ``n20'' and ``n20\_hiZ'' are magnetized everywhere in the gas ($\Theta_1<1,\,\Theta_2<1$), and indeed the dust-gas distributions are quite different to the uncharged case (``n20\_noZ''), with tighter coupling of the dust to the gas.  In contrast, the large grains ($\alpha=0.01$) are either unmagnetized (for standard grain charge, ``n20'') or in the ``mixed'' regime (for $10\times$ charge, ``n20\_hiZ''). In agreement with the theory, the standard-charge (``n20'') distribution looks similar to the uncharged case, while the ``n20\_hiZ'' distribution is similar at high gas densities (where the grains are unmagnetized) but exhibits stronger dust-gas coupling at low gas densities (where the grains are magnetized). For the parameters of this simulation (``n20\_hiZ'' $\alpha=0.01$), the critical gas density $n_{g,\mathrm{crit}}$ governing the change from magnetized to unmagnetized dust [see Eq.~\eqref{eq: critical density}]  is $n_{g,\mathrm{crit}}/\langle n_{\mathrm{gas}}\rangle\approx 75$, which is consistent with what is observed in Fig~\ref{fig:ndust.ngas.lorentz} (of course the change is gradual and the theory heuristic, so we should not expect obvious quantitative agreement).


The dust-to-gas ratio $\delta \equiv (n_{\rm dust}/n_{\rm gas}) / (\langle n_{\rm dust} \rangle / \langle n_{\rm gas} \rangle)$ (i.e., integrating out one dimension from Figs.~\ref{fig:ndust.ngas}-\ref{fig:ndust.ngas.lorentz}) is an interesting quantity  for both practical application to GMCs and for theory. We illustrate its distribution in Fig.~\ref{fig:vanilla} for each of the ``standard-charge'' simulations (``n1,'' ``n20,'' ``n100,'' ``n1000,'' and ``n10000''). As in \paperone, we find these are approximately log-normal, with power-law tails. More quantitatively, the dispersion of $\delta$ (denoted $\sigma_{\log \delta}$) is illustrated in Fig.~\ref{fig:2dparam} for all simulations (see also Table~\ref{tbl:sims}). There is clearly a strong increase in $\sigma_{\log \delta}$ with $\alpha$ --  i.e,. with larger $a_{d}$ and smaller $\langle n_{\rm gas}\rangle$ -- particularly  around $\alpha \sim 0.005-0.01$ where $\sigma_{\log \delta}$ increases from $\sim 0.05$ to  $\sim 0.35\,$dex. This is expected as grains transition from being tightly coupled to the gas for $\alpha \ll \mathcal{M}^{-2}$ to uniformly filling the box if $\alpha\gg1$ (see \paperone\ and Eq.~\ref{eqn: Ls}), and is a more quantitative illustration of the effects shown in Fig.~\ref{fig:ndust.ngas}. Also note that the ``floor'' at $\sigma \sim 0.015-0.02$ ($\alpha \lesssim 0.001$) is not real, but represents the limitations of our numerical method.\footnote{As discussed in \paperone\ (see also \citealt{genel:tracer.particle.method}), the mis-match between the EOM for grains, where gas quantities are interpolated to the exact grain location, and gas, where the fluxes are calculated from a Riemann problem and averaged over a finite volume, means there will inevitably be some small, purely numerical dust-to-gas fluctuations even when the two should be perfectly-coupled. There we showed $\alpha\lesssim 0.001$ hits this ``floor.''} Fig.~\ref{fig:2dparam} also serves to illustrate that the effects of dust magnetization  on the dispersion of $\delta$ are subdominant to its variation with $\alpha$, although the magnetized cases mostly show slightly lower  $\sigma_{\log \delta}$. In other words, the change to the dust-gas distribution with magnetization seen in Fig.~\ref{fig:ndust.ngas.lorentz} causes only a small modification to $\sigma_{\log \delta}$ in comparison to the variation with grain size. A A more detailed discussion of the non-magnetized scaling is given in \paperone.

It is helpful to examine the changes in dust-to-gas ratio distributions with individual parameters, which is done in Fig.~\ref{fig:machcomp.lorentzcomp}. The left panel shows the effects of Mach number $\mathcal{M}$ on the $\delta$  distribution with otherwise equal simulation parameters. We see that the effect on the $\delta$ distribution is minor, even though the logarithmic dispersion in the {\em gas} density in the higher-$\mathcal{M}$ run is significantly larger (by $\approx 0.2\,$dex, in agreement with the well-studied Mach number-density variance relation; \citealt{konstantin:mach.compressive.relation}). 
There is nonetheless some weak effect of $\mathcal{M}$ on $\delta$: at the lowest $\alpha$, the tails in $\delta$ are broadened (because dust in the lower-$n_{\rm gas}$ tails of the gas distribution is, locally, more weakly-coupled), while at large $\alpha$, the distribution actually becomes slightly more narrow (because the grains are already loosely-coupled, this moves the system more towards a ``uniformly mixed'' distribution).

In a similar vein, the right panel of Fig.~\ref{fig:machcomp.lorentzcomp} compares simulation with neutral grains (``n20\_noZ''), ``standard-charge'' grains (``n20''), and $10\times$ charged grains (``n20\_hiZ''), keeping all other parameters fixed. This is another way of examining the data shown in Fig.~\ref{fig:ndust.ngas.lorentz}. Similar to the discussion above, we see that the small grains ($\alpha=0.001$) in both the ``n20'' and ``n20\_hiZ'' simulations are quite different to the neutral grains (``n20\_noZ''), but similar to each other (aside from an increased dispersion in ``n20\_hiZ'', perhaps due to resonant acceleration; see below). In contrast, large grains ($\alpha=0.01$) are similar between the neutral and standard-charge grains\footnote{There is actually a slight enhancement in variance in ``n20'' compared to ``n20\_noZ'' for large grains. We speculate that this is because the lowest-gas density regions, which would before have completely de-coupled from the dust, have weak Lorentz coupling and induce some additional dust concentration.} (since these are unmagnetized), while the $10\times$ charged grains exhibit a substantial decrease in variance compared to the neutral grains because they are magnetized in low-gas-density regions (they are in the ``mixed'' regime).

Finally, it is worth commenting on an interesting feature of the $\delta$ distributions in Figs.~\ref{fig:vanilla} and \ref{fig:machcomp.lorentzcomp} -- the flat, high-$\delta$ tail that appears in some simulations (e.g., $\alpha=0.001$, ``n20''). A comparison to the parameters in Table~\ref{tbl:sims} shows that this exists only for those parameters at which the dust is magnetized ($\Theta_1<1,\;\Theta_2<1$), while the comparison to an equivalent neutral dust simulation in Fig.~\ref{fig:machcomp.lorentzcomp}(b) shows that it is related to the action of the Lorentz force (the tail appears only for charged grains and is stronger in ``n20\_hiZ'' compared to ``n20''). {We speculate that this effect may be related to resonant acceleration of dust grains, which can  
occur when multiples of the dust Larmor frequency match the turnover frequency  of the turbulence as seen by the dust \citep{2002ApJ...566L.105L,2003ApJ...592L..33Y,Yan:2004vh}. The turbulent magnetic 
field is then stationary in the dust frame and resonantly 
exchanges energy with the grains through Landau damping and cyclotron damping (as occurs for ions and electrons in weakly collisional  plasmas). The higher dust velocities could be particularly important
 for dust shattering and coagulation \citep{2003ApJ...592L..33Y}, but we leave further study of this interesting effect to
future work.}

\vspace{-0.5cm}
\section{Conclusions}
\label{sec:conclusions}

We study how charged dust grains behave in GMCs by running idealized simulations of isothermal, magnetized, super-sonic turbulence, with a population of dust grains of physically interesting sizes and realistic charge, which experience both drag and Lorentz forces from the gas. We argue that the dynamics are essentially determined by three dimensionless parameters, the turbulent Mach number $\mathcal{M}$, grain size parameter $\alpha \propto a_{d}$ (Eq.~\ref{eqn:grain.size}) and Lorentz parameter $\phi\propto Z_{d}/a_{d}^{2}$ (Eq.~\ref{eqn:lorentz.phys}). We show that, when $\mathcal{M}\gg1$, grain dynamics are strongly governed by the parameter $\alpha$. With small $\alpha\lesssim \mathcal{M}^{-2}$, dust moves tightly with the gas; with large $\alpha \gtrsim 1$, grains decouple from the gas and spread uniformly, while intermediate cases (expected for large grains in a wide range of typical GMCs) produce interesting local fluctuations in the dust-to-gas ratio $\delta$, with the logarithmic dispersion in $\delta$ increasing from $\sim 0.05 - 0.35\,$dex as $\alpha$ increases. At a given $\alpha$, we show that varying $\mathcal{M}$ (within the range expected in GMCs) has weak effects. Comparing simulations without Lorentz forces, we see the Lorentz forces produce a size-dependent effect: smaller grains (larger $\phi$) have their fluctuations suppressed with non-zero $\phi$, while larger grains show weak effects. This can be understood more quantitatively by considering the ratio of dust gyroradius $r_{\mathrm{gy,}d}$ to dust streaming length $L_{\mathrm{stream}}$, which we examine using the parameters $\Theta_1 = \alpha^{1/2} \mathcal{M}_A/\phi$ and $\Theta_2=  \mathcal{M}_A/\phi$ (for $\mathcal{M}_A\gtrsim 1$).  In general, $\Theta_1 \lesssim 1$ is required for appreciable effects on the dust clustering statistics, which implies that $r_{\mathrm{gy,}d}<L_{\mathrm{stream}}$ (at least at low densities), meaning the dust dynamics are significantly modified by the presence of the magnetic field. 

In \paperone\ (\S~4) we discuss implications of partial dust-gas coupling (and variations in local dust-to-gas ratios) for dust formation, extinction and observed dust clustering, cooling, and star formation. Because high-density regions can have enhanced/suppressed dust-to-gas ratios in large grains (which contain a large fraction of the metal mass), this can have interesting implications for stellar abundances. \citet{hopkins.conroy.2015:metal.poor.star.abundances.dust} use similar simulations, coupled to a specific dust chemistry model, to explore consequences for abundance patterns in metal-poor stars, and suggest that certain observed chemical signatures in these stars may demonstrate variable dust-to-gas ratios in their progenitor clouds. \citet{hopkins:totally.metal.stars} use a simple analytic model to further explore the consequences for stellar abundance variations across present-day star forming clouds. Taking observed scalings of GMC properties \citep[e.g.][]{bolatto:2008.gmc.properties} with size $\sim R_{\rm GMC}$, they show the critical parameter $\alpha/\mathcal{M}^{-2}$ should scale $\propto R_{\rm GMC}$ for grains of a fixed size. In physical terms, for sufficiently large clouds $\gtrsim 10-100\,{\rm pc}$ (for $0.1-1\,\mu m$ grains), $\alpha \gtrsim \mathcal{M}^{-2}$ and grain densities fluctuate on scales greater than the sonic length (the characteristic size of dense star-forming filaments and protostellar cores). All of this work, however, ignored Lorentz forces; our goal here was to explore how this might change the dynamics. Since we find the effects of Lorentz forces are sub-dominant to grain size variations, none of the key qualitative conclusions from these studies are altered. However, by further suppressing fluctuations in the small-grain regime (while having little effect for large grains), Lorentz forces will make the ``threshold'' effect above more dramatic (where fluctuations are unimportant below, but significant above, some characteristic grain/cloud size scale). 

A major caveat of this study is that we have considered only the cold, dense ISM in GMCs -- the values of $\alpha$ and $\phi$ here are appropriate when $T\lesssim 100\,$K. It is interesting to ask what happens to dust in the warm neutral and warm ionized medium, with $T\sim 10^{3}-10^{4}\,$K; since the equilibrium grain charge (and $\phi$) are expected to scale $\propto T$, we expect Lorentz forces to rapidly increase in importance. Unfortunately, the numerical method here (explicitly integrating the Lorentz forces) becomes unacceptably expensive for very large $\phi$ (as the Larmor frequency increases); we are working on a fully-implicit scheme for integrating the Lorentz term which will allow us to extend our simulations into this regime (also implicit schemes for including dust back-reaction; see \citealt{yang:integrator}). These simulations will also be interesting from a theoretical standpoint, allowing study of the magnetized, high-$\alpha$ region of parameter space that is absent from the simulation set presented in the current work (see, for example, Fig.~\ref{fig:2dparam}). In the mostly-ionized medium, we also need to account for Coulomb interactions, but these primarily manifest as a modest correction to the drag term \citep{draine.salpeter:ism.dust.dynamics} so their effect should be straightforward to understand and implement numerically.  Radiation pressure and dust collisional dynamics may also modify the conclusions here, especially in the most-dense regions where this is relevant for star formation, and we will explore this further in future work.

\vspace{-0.7cm}
\acknowledgments 
We thank the anonymous referee and Matthew Goodson for a number of useful comments and suggestions. Support for HL \&\ PFH was provided by NASA ATP Grant NNX14AH35G \&\ NSF Collaborative Research Grant \#1411920 and CAREER grant \#1455342. JS was funded in part by the Gordon and Betty Moore Foundation through Grant GBMF5076 to Lars Bildsten, Eliot Quataert and E. Sterl Phinney. Numerical calculations were run on Caltech cluster ``Zwicky'' (NSF MRI award \#PHY-0960291) \&\ XSEDE allocation TG-AST130039 supported by the NSF.\\ 
\vspace{-0.3cm}
\bibliography{ms2,jonos_extra_refs}

\end{document}